\documentclass[useAMS,usenatbib,letterpaper]{mn2e}   

\usepackage{graphicx}
\usepackage{natbib}
\title[TW~Hya: MOST \& ASAS observations in 2009]{Analysis of
variability of TW~Hya as observed by {\it MOST\/} and {\it ASAS\/} 
in 2009\thanks{Based on data from
the MOST satellite, a Canadian Space Agency mission, jointly operated
by Dynacon Inc., the University of Toronto Institute of Aerospace
Studies, and the University of British Columbia, with the assistance
of the University of Vienna, and from the All Sky Automated Survey
(ASAS) (conducted by the Warsaw University Observatory, Poland)
at the Las Campanas Observatory.}}

\author[M. Siwak et al.]
{Michal Siwak$^1$\thanks{E-mail: siwak@astro.utoronto.ca},
Slavek M.\ Rucinski$^1$,
Jaymie M.\ Matthews$^2$,
\newauthor
Grzegorz Pojma{\'n}ski$^3$,
Rainer Kuschnig$^{2,7}$,
David B.\ Guenther$^4$,
\newauthor
Anthony F.\ J.\ Moffat$^5$,
Dimitar Sasselov$^6$,
Werner W.\ Weiss$^7$\\
$^1$Department of Astronomy and Astrophysics,
University of Toronto, 50 St.\ George St., Toronto,
Ontario, M5S~3H4, Canada\\
$^2$Department of Physics \& Astronomy, University of
British Columbia, 6224 Agricultural Road, \\
Vancouver, B.C., V6T~1Z1, Canada\\
$^3$Astronomical Observatory, University of
Warsaw, Al. Ujazdowskie 4, 00-478 Warsaw, Poland\\
$^4$Institute for Computational Astrophysics,
Department of Astronomy and Physics,
Saint Marys University, \\  Halifax, N.S., B3H~3C3, Canada\\
$^5$D\'{e}partment de Physique and Centre de Recherche 
en Astrophysique du Qu\'{e}bec, Universit\'{e}
de Montr\'{e}al,\\ 
C.P.6128, Succursale: Centre-Ville,
Montr\'{e}al, QC, H3C~3J7\\
$^6$Harvard-Smithsonian Center for Astrophysics,
60 Garden Street, Cambridge, MA 02138, USA\\
$^7$Institut f\"{u}r Astronomie, Universit\"{a}t Wien,
T\"{u}rkenschanzstrasse 17, A-1180 Wien, Austria\\
}

\date{Accepted -- 2010 September 4.      Received -- 2010 September 2 ;      in original form -- 2010 August 1}

\pubyear{2010}

\begin{document}
\label{firstpage}
\maketitle

\begin{abstract}
As a continuation of our previous studies in 2007 and 2008,
new photometric observations of the T~Tauri star
TW~Hya obtained by the MOST satellite and 
the ASAS project over 40 days in 2009 with temporal resolution of 0.2 days are presented.
A wavelet analysis of the combined MOST--ASAS data
provides a rich picture of coherent, intermittent, variable-period
oscillations, similarly as discovered in the 2008 data.
The periods (1.3 -- 10 days) and systematic period
shortening on time scales of weeks 
can be interpreted within the model
of magneto-rotationally controlled accretion processes in the
inner accretion disk around the star. Within this model and
depending on the assumed visibility of plasma parcels 
causing the oscillations,
the observed shortest-period oscillation period may indicate
the stellar rotation period of 1.3 or 2.6 d, synchronized
with the disk at $4.5\,R_\odot$ or $7.1\,R_\odot$, respectively.

\end{abstract}


\section{Introduction}
\label{intro}

The data gathered by the MOST satellite in 2007
and 2008 for TW~Hya -- a T~Tauri-type star of spectral type K7Ve and V $\sim$11 mag
surrounded by an accretion disk --
were a subject of a Fourier and wavelet analysis by \citet{ruc08}.
The most interesting discovery resulted from the wavelet analysis, which
revealed oscillatory features systematically
shortening their periods on time scales
of tenths of days. With typical periods of 2 to 9 days,
a preliminary interpretation was through instability phenomena
within the innermost accretion disk.
In order to check whether such variations are
a constant feature of the star,
TW~Hya was observed by the MOST satellite for the third 
(and most likely last) time in 2009 (Section~\ref{obs}). 
Additionally, simultaneous {\it ASAS} observations
were obtained as close to the MOST observation in time as
possible.
In this paper we present an analysis of the new photometric data
(Section~\ref{analysis}) and propose a more physical interpretation
of the features discovered in the wavelet spectrum (Section~\ref{interp}).

\begin{figure*}
\includegraphics[height=175mm,angle=-90]{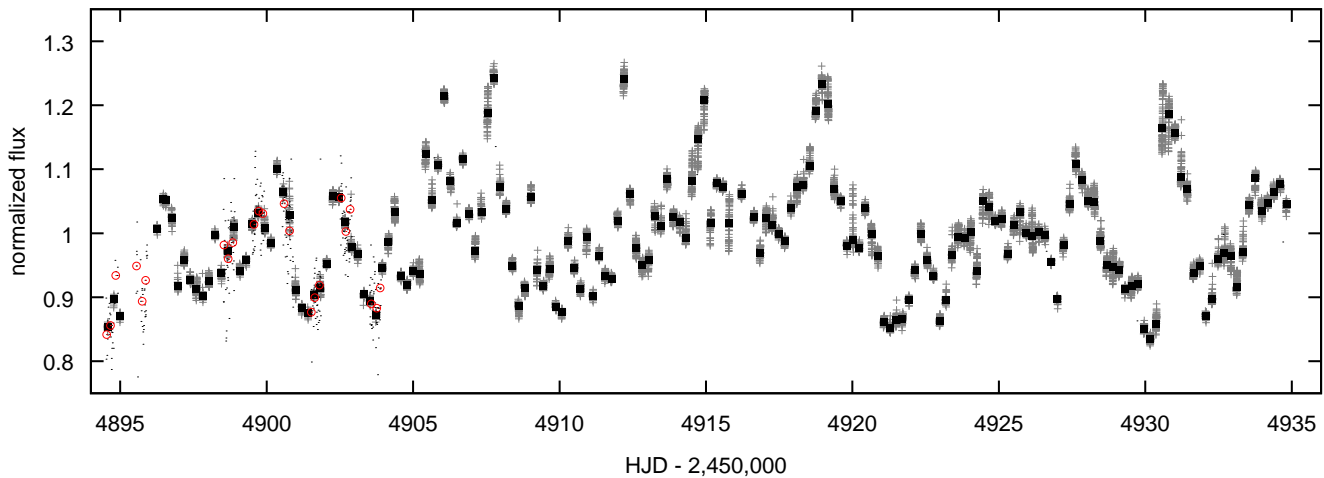}
\caption{The 2009 light curve of TW~Hya in flux units,
scaled (separately for each source) to unity at the mean brightness level.
The original MOST data are presented as crosses and
their respective median-orbital values
are shown as black squares. The ASAS observations obtained
at $hjd=4894.5-4903.9$ in the Johnson $V$ filter
with 10~minute temporal resolution, expressed in flux units
assuming mean magnitude $\bar{V} = 11.065$, are shown
as small dots; their mean values, calculated for intervals
of 0.2114~day, corresponding to the MOST data collection
rate of 3 satellite orbits, are shown as circles.
Note the good match between the MOST and ASAS data sets where
they overlap during the first nine days of the MOST observations.}
\label{Fig.1}
\end{figure*}

\section{Observations and data reduction}
\label{obs}

The optical system of the MOST satellite consists
of a Rumak-Maksutov f/6 15~cm reflecting telescope.
The custom broad-band filter covers the spectral range of
380 -- 700~nm with effective wavelength falling close
to Johnson's $V$ band.
The pre-launch characteristics of the mission are described
by \citet{WM2003} and the initial post-launch performance
by \citet{M2004}.

TW~Hya was observed between March 3 and April 12, 2009,
corresponding to the $hjd$ range
of $4894.53 - 4934.83$ \footnote{Throughout this
paper: $hjd \equiv HJD-2,450,000$}.
The data were collected during the low
stray-light segment, lasting typically 30~mins, of 
(almost) every third satellite orbit over 40.3~days.
The direct-imaging mode of the satellite was
used \citep{WM2003} with individual exposures of 30~sec.

\begin{figure}
\includegraphics[width=61mm,angle=-90]{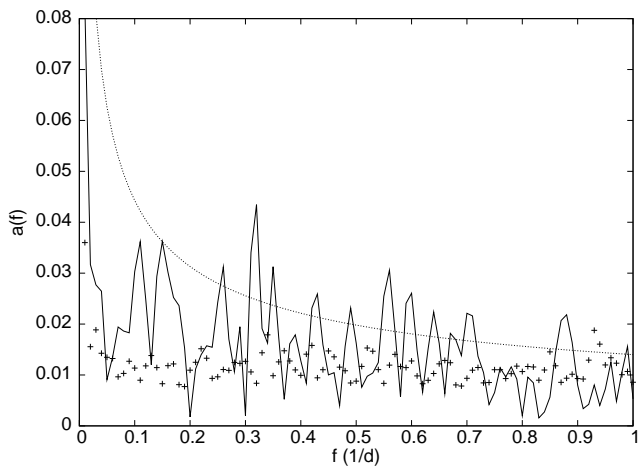}
\caption{The frequency spectrum of TW~Hya obtained from the combined,
time-averaged 2009 MOST (185 points) and ASAS $V$ filter (23 points) data.
No statistically significant peaks are present above the frequency of 1~c/d.
The dotted line shows the arbitrarily scaled $1/\sqrt{f}$ dependence, while
the crosses represent errors of amplitudes $a_{i}(f)$.}
\label{Fig.2}
\end{figure}

The data were reduced in the same way as described by \citet{siwak}.
Small linear correlations between the stellar flux
and the sky background level within individual orbits
(most probably caused by a small
photometric nonlinearity of the electronic system) were also removed.
In the last step the data were de-trended by means
of a low-order polynomial fitted to the light curve of
a constant star, simultaneously observed with the main target.
As a result, we obtained a very good quality light curve
with median error of a single data point of about 0.009~mag.
The systematic errors related with the uncertainty of the nonlinearity
corrections do not exceed 0.005 mag and are inconsequential
for our investigation.

\begin{figure*}
\includegraphics[width=120mm]{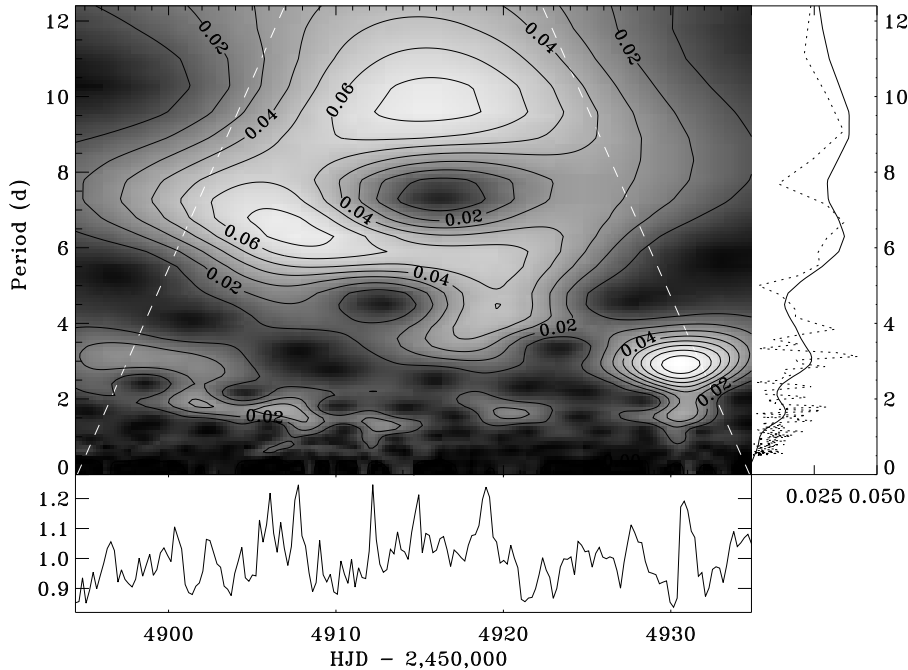}
\caption{The wavelet transform of the 2009 TW~Hya MOST data supplemented by
the ASAS data, as explained in the text.
The amplitude power of the transform is expressed by the grey scale intensity plus contours with arbitrary units.
Edge effects are present outside the white broken lines; they do not
affect our conclusions.
At the bottom, the combined MOST and ASAS-$V$ light curve, re-sampled
into a uniformly distributed time-grid with 0.2114~d spacing, is shown;
at the right a comparison is given of the restored (via horizontal
averaging) frequency spectrum (the continuous line)
with the power spectrum corresponding to
the amplitude spectrum in Fig.~\ref{Fig.2} (the dotted line).
}
\label{Fig.3}
\end{figure*}

In parallel with the MOST observations, TW~Hya was also monitored
by the {\it All Sky Automated Survey} (ASAS) \citep{pojmanski} 
in Johnson V filter with temporal sampling of about 10~min.
Due to deterioration of the CCD chip used for $V$-band observations
over the last years, the individual data points have rather
low accuracy with a median error of 0.045~mag.
The mean observed magnitude, $\bar{V}=11.065$, was
used for conversion of the ASAS data
to normalized flux units.
Similarly as in 2008 \citep{ruc08}, the 2009 data show
good correspondence between MOST and the ASAS $V$-band.
For comparison with the MOST observations,
the ASAS $V$-band data points were averaged using the sampling
rate of the MOST data of 0.21142~d.

In Figure~\ref{Fig.1} we present the original and the averaged data points
of TW~Hya obtained by the MOST satellite and ASAS in Johnson $V$ filter
in normalized flux units.
The median error of the time-averaged MOST data in flux units is equal
to 0.0066.
The time averaged ASAS $V$-band data have an order of
magnitude worse accuracy, of about 0.041 flux units.
For both datasets, the errors include 
the genuine, time-averaged, short-timescale
physical variability of TW~Hya observed within individual {\it MOST} 
orbits; thus, the rapid variability of the star was lost in the averaging 
process, similarly as in the previous analysis of the 2007 and 2008 
data \citep{ruc08}.

\section{Analysis of {\it MOST} 2009 data}
\label{analysis}

\subsection{Fourier transform}
\label{ft2009}

We performed a Fourier analysis of the 2009 data 
as in \citet{ruc08}
(see Section~4.2 of that paper for details).
The frequency spectrum was obtained using the light curve 
formed from the averaged MOST and ASAS $V$-band data,
represented by squares and circles, respectively, in Figure~\ref{Fig.1}.
The 0.2114~d sampling rate ($3\times$ the MOST orbit duration)
precludes analysis of frequencies higher than 2.35~c/d.
In Figure~\ref{Fig.2}, we show only the frequency range below 1~c/d
because there are no statistically significant peaks
at higher frequencies.
We note that \citet{gunther} did not find any sub-minute
periodicities in their high-speed time-series photometry
of TW~Hya obtained with the SALT telescope.

Because the temporal resolution of the 2009 data was
$3\times$ poorer than that of the 2008 data,
the amplitude errors are now larger than before, with
median value of the error 0.012 of the mean flux.
As observed before, the
peak amplitudes $a(f)$ seems to scale as flickering noise
($a(f) \sim 1/\sqrt{f}$),
but due to the larger errors, this tendency is not
as obvious as in the 2008 data \citep{ruc08}.

\subsection{Wavelet analysis}
\label{wav2009}

The original time-distribution of averaged data points differs slightly from
the uniformity required by the wavelet technique. Therefore, we interpolated
the data into a uniform time grid using splines.
Except for the orbital-averaged MOST data points, we also included
the first five average ASAS points obtained in $V$-band before and during
the gap in the MOST data which occurred within
$hjd \approx 4895-4896.5$.
The remaining ASAS points were not used because the ASAS data were
significantly less accurate and were used here only to assure uniformity
of the main {\it MOST} sequence for the wavelet analysis.

Figure~\ref{Fig.3} shows the result of the wavelet transform of
191 data points uniformly distributed in time at the spacing of 0.2114~d.
As in the case of the 2008 data, the Morlet-6 wavelet provided the best match
between the time-integrated power spectrum and original
frequency spectrum, as is visible in the right panel
of Fig.~\ref{Fig.3}. We stress that only for the Morlet-6 wavelet
do we obtain such a nearly-perfect match of both plots;
as described in \citet{ruc08}
Morlet transforms of other orders result in systematic
differences in the period scale.

\begin{table}
\caption{Numerical values extracted from the wavelet spectrum and related
physical parameters computed assuming that the features observed in
the spectrum come from a Keplerian disk around the central star.\newline
The features marked with an asterisk $^*$ are for the 
data in \citet{ruc08}.}
\centering{
\begin{tabular}{l c c c r r}
 \hline\hline
Feature & $hjd$ & $P\,[d]$ & {\it{\.P}} & $a_1\,[R_{\odot}]$ & $a_2\,[R_{\odot}]$ \\ \hline\hline
 1 - start     & 4901 & 1.9 &           & 5.8  & 9.2  \\
 1 - end       & 4913 & 1.3 & -0.049(2) & {\bf 4.5}  & {\bf 7.1}  \\ \hline
 2 - start     & 4898 & 3.1 &           & 8.0  & 12.7 \\
 2 - end       & 4922 & 1.6 & -0.065(1) & 5.2  & 8.2  \\ \hline
 3 - start     & 4902 & 7.2 &           & 14.1 & 22.3 \\
 3 - end       & 4931 & 3.0 & -0.151(3) & 7.8  & 12.5 \\ \hline
 4 - start     & 4914 & 9.8 &  ---      & 17.3 & 27.4 \\ \hline
 5$^*$ - start & 4533 & 6.0 &           & 12.5 & 19.8 \\
 5$^*$ - end   & 4553 & 3.0 & -0.143(2) & 7.8  & 12.5 \\ \hline
 6$^*$ - start & 4556 & 7.5 &  ---      & 14.5 & 22.9 \\ \hline\hline
\end{tabular}
\label{Tab.1}
}
\end{table}

\section{Discussion}
\label{interp}

The most important result of this study is that the general behaviour
of the Morlet wavelet transform in the {\it time -- period}
plane is similar to that obtained and described by \citet{ruc08}
for the 2008 observing season.
The systematically migrating oscillatory features that we saw
in 2008 were not an unusual phenomenon limited to this particular point in time;
apparently, TW~Hya shows this phenomenon all or most of the time.
The new 2009 data not only confirm the 
direct visibility of the features drifting to shorter
periods, but this time we could identify as many as four of them.
Their approximate  moments of appearance
and disappearance (in {\it hjd}), 
the periods {\it P} observed at the beginning and at the end
of each oscillation packet and the 
respective Keplerian radii $a_1$ calculated
assuming $M_{*}=0.72~M_{\odot}$ and $R_{*}=1~R_{\odot}$ are presented
in Table~\ref{Tab.1}; they are ordered in their start periods.
The data in Table~\ref{Tab.1} for features no. 1--3 were
obtained by tracing local maxima along the patterns
in Figure~\ref{Fig.3}; similarly we used figure~8
of \citet{ruc08} to trace feature no.5* identified
in the 2008 data.
We did not investigate features nos. 4 and 6*: the former
had a long period of almost 10~d, so that
the 2009 run was too short to fully follow its evolution while
the latter feature (from the 2008 data) occurred too close
to the end of that observing run.

\begin{figure}
\includegraphics[width=84mm]{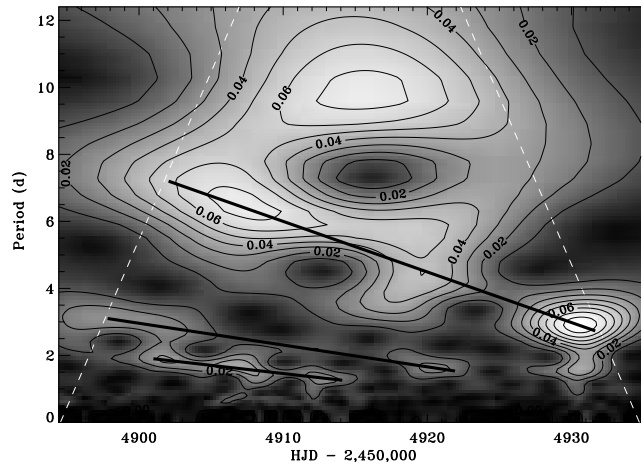}
\caption{The linear fits (thick black lines) to the local maxima
of features 1 -- 3 visible in Fig.~\ref{Fig.3}.
Similar fits were made for the features observed in
2008 of which only feature no.5* could provide sufficiently well defined
results.}
\label{Fig.4}
\end{figure}

Within the accuracy of locating the local peaks, only the lowest-order
fits by straight lines (see Figure~\ref{Fig.4}) gave a
sufficiently significant description of their temporal
evolution described by their time derivatives 
{\it{\.P}},
as given in Table~\ref{Tab.1}. The fits were weighted by their local peak
intensities.
We note that the rate of the period change appears to be faster for
longer periods.

In the now most popular interpretation, originally developed for accreting
neutron stars by \citet{ghosh77,ghosh79a,ghosh79b} and further applied
to classical T~Tau-type stars (CTTS) \citep{konigl91, cameron93, hartmann94,
shu94, muzerolle98, muzerolle01}
magnetic loops originating on the stellar surface connect
to the accretion disk outside the co-rotation radius to avoid angular
momentum transfer from the disk.
The model of magnetically controlled accretion explains observations
of many CTTS, e.g. AA~Tau \citep{bouvier07} and BP~Tau \citep{donati}, 
as well as some young brown dwarfs, e.g., 2MASSW J1207334-393254 in 
the TW Hya association \citep{scholz}.

As proposed before \citep{ruc08}, we think that
the observed light oscillations
are produced in one or several regions in the inner accretion disk due
to instabilities in the disk plasma. They may be caused by magnetic
field threading of the disk near its inner edge via Kelvin-Helmholtz
instability or by combined turbulent diffusion and reconnection.
\citet{ghosh79a,ghosh79b} found that such an interaction would
produce a broad transition zone between the unperturbed 
Keplerian disk and the co-rotating magnetosphere 
outside which considerable departures
from Keplerian motion are expected.
Their theoretical expectation for the {\it outer transition zone\/} seems
to be confirmed by \citet{calvet} (their Sec.~4.3), who found that at
a distance of 1~AU, gas in the TW~Hya disk probably drifts slowly
inward nearly following Keplerian orbits. 
In this picture, the oscillations observed by MOST
may be explained by orbital motion of optically
thick condensations produced by {\it magnetic loop -- disk}
interactions in the {\it inner transition zone\/} 
within more spatially confined regions, bright enough 
to produce the observed, large amplitudes of light variations.
Once initiated at respective Keplerian radii $a_1$ 
(see Table~\ref{Tab.1}),
they would then rapidly propagate toward the co-rotation and 
truncation radii, a property which could explain the high absolute 
values of {\it{\.P}}. 

Although there may be doubts if the stellar rotation -- disk revolution
synchronization still operates in TW~Hya (see below),
one could this way estimate the period of rotation of the star.
The shortest period observed in the MOST data 
(the feature no.1 at the end of its evolution at $hjd=4913$)  
is $1.3 \pm 0.1$~d; admittedly, this period is poorly defined
at the $2.2\sigma$ significance level. 
This value is almost identical to the shortest period
among many periods previously claimed for
TW~Hya of 1.28~d \citep{hk88}
(see sect.~2 of \citet{ruc08} for a summary of this topic).
Similar values ($1.4 \pm 0.1$~d and $1.55 \pm 0.10$~d)
have been obtained by \citet{alencar} in an analysis
of veiling measurements, but the authors disregarded these
values paying more attention to other detected periodicities,
i.e.\ 2.9 and 4.5~d.
However, as obtained by \citet{wilner} from 7~mm {\it VLA}
interferometric observations, the outer accretion disk
is oriented almost face-on. This means -- 
for the inner accretion disk being co-planar with its outer parts
-- that each hypothetical condensation
causing observed light oscillations is seen twice 
during each orbital evolution.
Then, the observed periods given in Table~\ref{Tab.1} would
in fact represent half-values of the real orbital periods.
Accordingly, we re-calculated the Keplerian distances assuming 
doubled values of the observed periods, which are
given as $a_2$ in the last column of Table~\ref{Tab.1}.
In this scenario, the value of $2.6\pm0.2$~d rather than $1.3\pm0.1$~d,
would correspond to the pure
rotational period of the central star itself.

Several semi-direct estimates of the inner disk radius are available
from the literature: \citet{weinberger},
using spatially resolved {\it HST} and {\it Keck}  infrared images
and spectra, modeled assuming different dust grain sizes,
obtained $10.7 \pm 2.1~R_\odot$ for the inner disk radius.
\citet{eisner}, 
using  spatially resolved near $K$-band interferometric {\it Keck}
observations and a model utilizing submicron-sized dust grains,
obtained $12.9\pm2.1~R_{\odot}$.
Contrary to \citet{weinberger}, their second
model with large dust grains
gave a smaller radius of $4.3 \pm 2.1~R_{\odot}$,
but much poorer fit to all the data.
We note that only if we assume the doubled orbital period
do we obtain an inner disk radius of $7.1\pm0.4~R_{\odot}$,
which would be commensurate with these estimates
(Table~\ref{Tab.1}).

In spite of its conceptual appeal, the coupling between
stellar and disk rotation offered by the magnetically controlled
accretion model may no longer be valid for TW~Hya.
Although TW~Hya possesses properties of a young (1--2~Myr old)
CTTS, it is part of an evolved ($\simeq 12 \pm 8$~Myr old)
TWA association, as obtained from lithium depletion estimates
of \citet{mentuch}.
The accretion disk itself also shows indications
of being in advanced stages of dust evolution: as previously
noted, \citet{weinberger} and \citet{calvet} needed large-size
dust grains to fit the observed spectral energy distribution;
the latter authors discovered also a developing gap in the disk
at the distance of 4~AU from the central star. 
From an analysis of several young stellar associations,
\citet{rayjay06} found that accretion disks in T~Tau-type
stars appear to cease before the age of about 10~Myr:
among 32 members of the TWA association, only TW~Hya and TWA~3A
show signs of ongoing accretion and seem to be slow rotators.
We note that the small value of $v\,\sin i = 10.6~km/s$ determined by
the authors for TW~Hya should not be taken to
indicate a long rotational period of the star, 
but is rather a result of the low inclination angle of its rotation
axis \citep{wilner}.

Because of the pole-on geometry of TW~Hya, \citet{alencar} conclude 
that it is rather unlikely that 
the observed light variations are caused by rotation of hot spots on 
the star photosphere, especially in the case when magnetic poles seems 
to be aligned with the stellar rotation axis (see Section~4 of their 
paper for detailed explanation). These considerations may also
apply to the theoretical prediction of regularities
in the geometrical spot distribution as well as quasi-periodic light 
oscillations caused by hot spots on the stellar photosphere 
\citep{kulkarni,romanova}.
Although the oscillations may have similar characteristics
to light variations observed by {\it MOST}, a long rotational 
period of TW~Hya (of order 10~days) would be 
required to be explained by such a mechanism. We note that 
neither the 2008 nor the 2009 {\it MOST} observations revealed any 
common for both runs, long-period (8-15~d) wavelet-transform features
which could be attributed to rotational period of the star itself. 
Although this result may be caused by the limitation of a single 
{\it MOST} monitoring run to about 45~day, there are no indications of
such a long rotational period in the Fourier spectrum of the long
series of {\it ASAS\/} \citep{ruc08} observations.

\section{Conclusions}

The wavelet transform analysis of the 2009 data obtained  
by the MOST satellite and the ASAS project
confirms the essential findings of \citet{ruc08}: The
general red-noise nature of the Fourier spectrum of TW~Hya is
rather complex in that it corresponds to several oscillation packets
with their frequencies increasing in time and with larger
amplitudes for lower frequencies.
The respective orbital radii can be calculated assuming that periods
of individual oscillation packets are related to Keplerian motion
of optically thick plasma condensations in the inner
accretion disk. 
The observed periods are consistent with their origins
in the inner accretion disk, several stellar radii outside
of the co-rotation and truncation radii.
Noting the pole-on geometry of the accretion disk for TW~Hya, 
consideration should be given to the possibility 
that the observed periods represent
the respective halves of the true orbital periods.
Assuming that we detected the shortest possible period in the Morlet wavelet
transform of 1.3~d and that the inner boundary of the disk co-rotates with
the star, the co-rotation radius for 
TW~Hya would be about $4.5\pm0.2~R_{\odot}$
or $7.1\pm0.4~R_{\odot}$ depending on the assumed period
($1.3\pm0.1$ or $2.6\pm0.2$~d, respectively); 
only for the longer period is the inner disk size 
in accordance with the results of \citet{weinberger} and \citet{eisner},
obtained from spatially-resolved infrared images 
and interferometric observations.
Determination of the rotational period of the star
is crucially important both for the inner accretion disk 
considerations as well as for mechanisms investigated theoretically
by \citet{kulkarni} and \citet{romanova} which could
explain light variations through (possibly time-evolving) geometrical
distribution of hot spots and/or quasi-periodic oscillations in this 
region.

\section*{Acknowledgments}

MS acknowledges the Canadian Space Agency Post-Doctoral position
grant to SMR within the framework of the Space Science Enhancement 
Program and thanks Prof.\ Ray Jayawardhana for advice and comments.\newline 
The Natural Sciences and Engineering Research Council of
Canada supports the research of DBG, JMM, AFJM, and SMR.
Additional support for AFJM comes from FQRNT (Qu\'ebec).
RK is supported by the Canadian
Space Agency and WWW is supported by the Austrian Space
Agency and the Austrian Science Fund.

Special thanks are due to the referee, Dr. W. Herbst, for
very useful comments and suggestions.

\end{document}